\documentclass[twocolumn]{aastex63}%

\usepackage{amssymb,amsmath}
\usepackage{graphicx}
\usepackage{xcolor,natbib}
\usepackage{multirow}
\usepackage[T1]{fontenc}
\usepackage{ae,aecompl}
\usepackage{newtxtext, newtxmath}
\usepackage{float}
\usepackage{subfigure}
\usepackage{longtable}
\usepackage{enumitem}
\usepackage{longtable,listings}
\usepackage[flushleft]{threeparttable}
\usepackage{parcolumns}
\usepackage{hyperref}

\def\obj{SDSS J1347+1217}
\def\n{[N~{\sc ii}]$\lambda6584$\AA}

\def\s{[S~{\sc ii}]$\lambda6716$\AA}

\def\o{[O~{\sc i}]$\lambda6300$\AA}

\shorttitle{SDSS J1347+1217}
\shortauthors{Cheng et al.}

\begin{document}

\title{AGN-Driven Biconical Outflows as the Origin of the Double-Peaked [O~{\sc iii}] doublet in SDSS J134733.36+121724.27}

\author{PeiZhen Cheng}
\affiliation{Guangxi Key Laboratory for Relativistic Astrophysics, School of Physical Science and Technology, GuangXi University, No. 100, Daxue Road, Nanning, 530004, P. R. China}

\author{XingQian Chen}
\affiliation{Guangxi Key Laboratory for Relativistic Astrophysics, School of Physical Science and Technology, GuangXi University, No. 100, Daxue Road, Nanning, 530004, P. R. China}

\author{GuiLin Liao}
\affiliation{Guangxi Key Laboratory for Relativistic Astrophysics, School of Physical Science and Technology, GuangXi University, No. 100, Daxue Road, Nanning, 530004, P. R. China}

\author{Qi Zheng}
\affiliation{Guangxi Key Laboratory for Relativistic Astrophysics, School of Physical Science and Technology, GuangXi University, No. 100, Daxue Road, Nanning, 530004, P. R. China}

\author{Ying Gu}
\affiliation{Guangxi Key Laboratory for Relativistic Astrophysics, School of Physical Science and Technology, GuangXi University, No. 100, Daxue Road, Nanning, 530004, P. R. China}

\author{MuLin Chen}
\affiliation{Guangxi Key Laboratory for Relativistic Astrophysics, School of Physical Science and Technology, GuangXi University, No. 100, Daxue Road, Nanning, 530004, P. R. China}

\author{XueGuang Zhang$^{*}$}
\affiliation{Guangxi Key Laboratory for Relativistic Astrophysics, School of Physical Science and Technology, GuangXi University, No. 100, Daxue Road, Nanning, 530004, P. R. China}

\correspondingauthor{XueGuang Zhang}
\email{xgzhang@gxu.edu.cn}

\begin{abstract}
In this manuscript, we recheck the spectroscopic properties of SDSS J134733.36+121724.27 (4C+12.50), confirming the presence of the double-peaked [O~{\sc iii}]$\lambda\lambda4959,5007$\AA\ doublet and a broad H$\alpha$.
The former likely results from AGN-driven biconical outflows, while the absence of a broad H$\beta$ supports a classification of the source as a Type-1.9 AGN.
We analyze its high-quality Sloan Digital Sky Survey (SDSS) optical spectrum after robustly subtracting host galaxy and AGN continuum contributions through a simple stellar population fitting method employing 39 templates and a power-law continuum.
Each narrow line of the [O~{\sc iii}]$\lambda\lambda4959,5007$\AA\ doublet is better described by two Gaussian components (blue-shifted and red-shifted) than by a single Gaussian, as confirmed by the F-test.
Broad components are included for both H$\alpha$ and H$\beta$, but only H$\alpha$ reveals a significant detection, further supported by a comparison between the SDSS spectrum and that previously reported. 
These results support that the object is highly consistent with a Type-1.9 AGN classification, and the double-peaked [O~{\sc iii}] profiles are most likely produced by AGN-driven biconical outflows rather than by a rotating narrow-line region or a dual AGN merger system.
Additional observations are still needed to strengthen these conclusions.
\end{abstract}


\section{Introduction}

SDSS J134733.36+121724.27 (also known as 4C+12.50 or PKS 1345+12, hereafter as \obj) with a redshift of $z\sim0.12$ has been the subject of extensive optical and near-infrared spectroscopic studies for more than four decades. 
In an early study, based on low-resolution spectroscopy with resolutions of about 10\AA, \citet{1977ApJ...215..446G} reported that the [O~{\sc iii}]$\lambda\lambda4959,5007$\AA\ emission lines exhibited broad and single-peaked profiles with pronounced blue-ward wings. 
This was accompanied by redshift discrepancies among emission lines, including [Ne~{\sc iii}]$\lambda\lambda3869,3967$\AA, the Balmer series, and low-ionization species such as [O~{\sc i}], [N~{\sc ii}], and [S~{\sc ii}]. 
Subsequently, \citet{1986Natur.321..750G} revealed the double-peaked structures of [O~{\sc iii}]$\lambda\lambda4959,5007$\AA\ doublet through a higher quality optical spectrum with resolutions of about 1\AA, thereby revising the earlier single-peaked interpretation (see their Fig.~2). 
In the near-infrared, \citet{1997ApJ...484...92V} reported the broad Pa$\alpha$ emission line for the first time, suggesting the presence of a broad-line region (BLR) in \obj. 
\citet{1999MNRAS.305..829I} further confirmed the line luminosity of the broad H$\alpha$ line.
Based on observations from the San Pedro Mártir Observatory in Mexico, \citet{2012RMxAA..48....9T} reported a possible broad H$\beta$ component.

It has been pointed out that mechanisms such as dual AGN merger systems, rotating narrow-line regions (NLRs), and AGN-driven biconical outflows may all lead to either double-peaked narrow emission lines or very extended components in narrow emission lines \citep{2004ApJ...604L..33Z, 2009ApJ...705L..20X, 2009ApJ...705L..76W, 2011ApJ...738L...2M, 2012ApJS..201...31G, 2013ApJ...762..110L, 2016ApJ...832...67N, 2018ApJ...867...66C, 2020ApJ...904...23K, 2024MNRAS.531L..76Z}.
Therefore, systematically investigating the possible physical connections between the emission line properties of \obj\ and its distinctive morphological features and associated outflows is of significant scientific interest and constitutes the central objective of this manuscript.

In addition to previously reported optical spectroscopic results in the literature, the high-quality optical spectrum obtained by the Sloan Digital Sky Survey (SDSS) on May 16, 2004, provided critical data for the reclassification and profile analysis of \obj.
However, significant discrepancies remain among studies regarding the interpretation of its spectroscopic features and the AGN classification.
\citet{2015ApJS..219....1O} classified it as a Type-1 AGN with a BLR, due to the prominent broad H$\alpha$. 
In contrast, \citet{2018ApJ...859..116K} treated it as a Type-2 quasar with strong [O~{\sc iii}] emissions but without broad Balmer emission lines. 
\citet{2022MNRAS.511..214N} also treated \obj\ as a Type-2 AGN, as shown without considering the effects of central AGN continuum emissions on the best descriptions of host galaxy properties by the simple stellar population (SSP) method. 
\citet{2023MNRAS.518..130S} further supported \obj\ as a Type-2 AGN and also showed that there are no apparent double-peaked profiles in the [O~{\sc iii}]$\lambda\lambda4959,5007$\AA\ doublet of \obj. 
Therefore, until now, there have been contrary conclusions on the classifications of \obj\ and also on the emission profiles of the [O~{\sc iii}]$\lambda\lambda4959,5007$\AA\ doublet, highlighting the need to recheck the spectroscopic properties of \obj.

Moreover, beyond the discussions of its spectroscopic properties, the physical properties of this object have been investigated from various angles. 
\obj\ is widely recognized as a radio-loud object exhibiting a characteristic S-shaped double radio jet structure \citep{2003ApJ...584..135L}. 
As early as the study by \citet{1986Natur.321..750G}, this object was identified as a galaxy merger system with two nuclei separated by 1.8\arcsec, exhibiting complex morphological features in the optical band. 
Similar optical morphological properties and the corresponding discussions can also be found in \citet{1993ApJS...88....1S}, \citet{2000AJ....120.2284A}, \citet{2016A&A...596A..19E}, and \citet{2022MNRAS.513.4770F}. 
Meanwhile, based on $^{12}$CO observations, \citet{2012A&A...541L...7D} reported the presence of prominent molecular gas outflows in \obj, with the outflow properties further quantified by \citet{2022MNRAS.515.1705Z}.

In this manuscript, we report a rare AGN whose emission lines all exhibit systematic blueshifts, with the broad H$\alpha$ component showing a significantly larger shifted velocity than the narrow lines. 
The high-quality SDSS spectrum of \obj\ is used to investigate its emission line properties, which help answer the following questions. 
\begin{enumerate}
  \item Are there double-peaked profiles in the [O~{\sc iii}]$\lambda\lambda4959,5007$\AA\ doublet of \obj? 
  \item Are there broad Balmer lines in the spectrum of \obj? 
  \item If there are double-peaked [O~{\sc iii}] profiles, what is the physical origin: due to dual AGN merger systems, rotating NLRs, or AGN-driven biconical outflows? 
\end{enumerate}
Then, this manuscript is organized as follows. 
Main results on the SDSS spectrum of \obj\ are shown in Section \ref{main result}. 
Necessary discussions, especially on the three questions, are provided in Section \ref{discussions}. 
Our main summary and conclusions are given in Section \ref{summary}.
In this manuscript, the cosmological parameters have been adopted as $H_{0} = 70{\rm km\cdot s}^{-1}{\rm Mpc}^{-1}$, $\Omega_{\Lambda} = 0.7$, and $\Omega_{\rm m} = 0.3$.

\section{Spectroscopic results of \obj\ in SDSS} \label{main result}

\begin{figure}
\centering\includegraphics[width = 9cm,height=5cm]{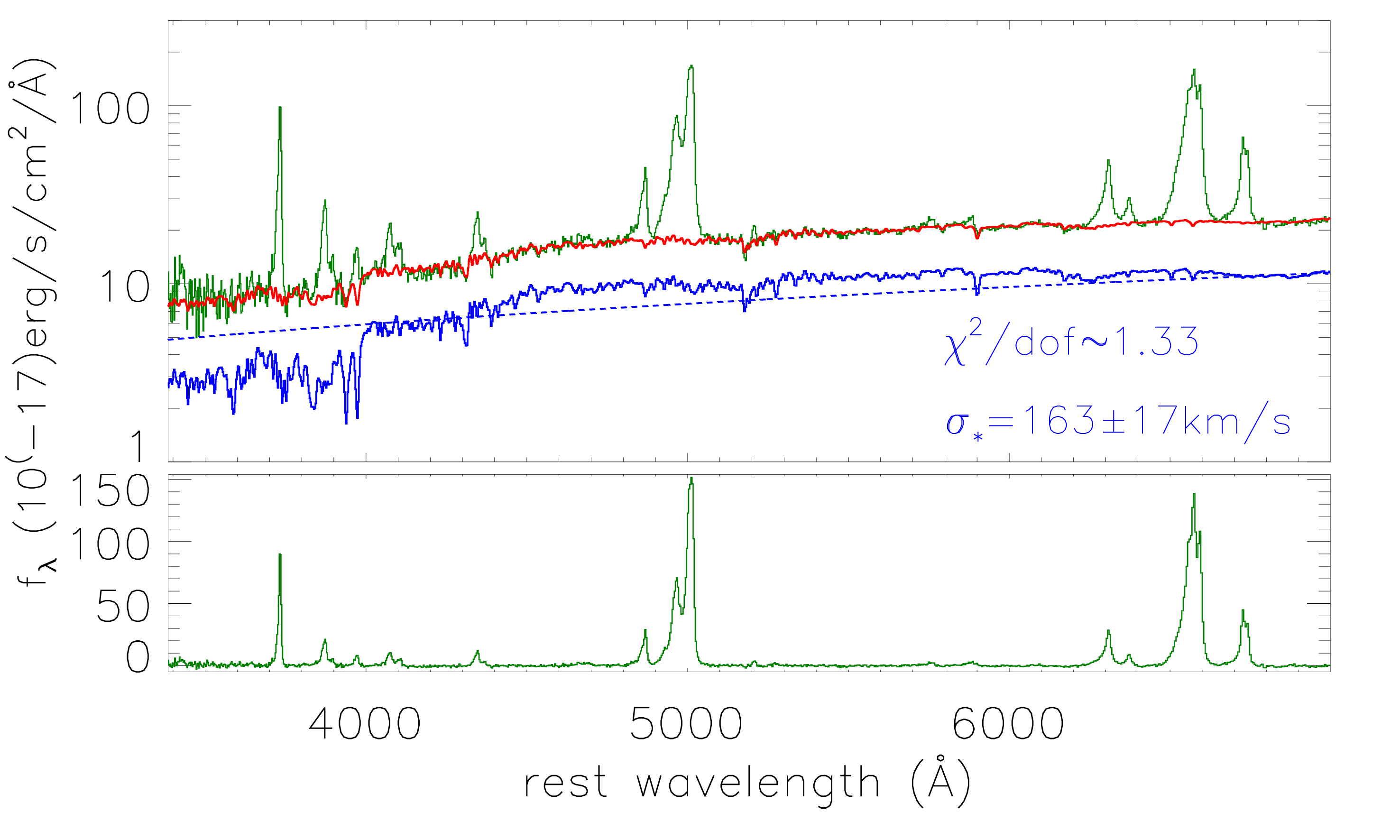}
\caption{The best descriptions of the host galaxy contributions in the SDSS spectrum of \obj. 
In the top panel, in order to show clearer absorption features, the y-axis is plotted in logarithmic coordinates. 
The solid dark green line shows the SDSS spectrum, the solid blue line shows the SSP method-determined host galaxy contributions, 
the dashed blue line shows the determined AGN continuum emissions, and the solid red line shows the sum of the host galaxy contributions and the AGN continuum emissions. 
In the bottom panel, the line spectrum is shown as the SDSS spectrum minus the host galaxy contributions and the AGN continuum emissions.}
\label{spec} 
\end{figure}

The SDSS spectrum of \obj\ has been collected from SDSS DR16 (Data Release 16; \citealt{2020ApJS..249....3A}), and is shown in the top panel of Fig.~\ref{spec}. 
In order to determine more reliable parameters of the emission lines in the SDSS spectrum, the following two steps are applied.

For the first step, the commonly accepted SSP method has been applied to determine the host galaxy contributions in the SDSS spectrum of \obj.
Similar as what we have recently done in \citet{2023ApJ...948...68Z, 2023ApJS..267...36Z, 2023MNRAS.519.4461Z, 2024ApJ...960..108Z, 2024ApJ...961...82Z, 2024ApJ...964..141Z, 2024MNRAS.529...41Z}, the host galaxy contributions are estimated using 39 SSP templates from \citet{2003MNRAS.344.1000B}, which have been widely applied to characterize almost all SDSS galaxies \citep{2003MNRAS.344.1000B, 2003MNRAS.346.1055K}. 
More detailed descriptions of the SSP method can be found in \citet{2003MNRAS.344.1000B}, \citet{2003MNRAS.346.1055K}, \citet{2005MNRAS.358..363C}, and \citet{2017MNRAS.466..798C}. 
Given the presence of broad H$\alpha$ in the SDSS spectrum of \obj, a power-law component has been applied to describe the AGN continuum emissions.
The presence of the broad H$\alpha$ component and the justification for adopting a power-law continuum will be further discussed in detail later in this manuscript.

\begin{table*}
\footnotesize
\caption{Line parameters of the emission lines in \obj}
\centering
\begin{tabular}{lcccc|lcccc}
\hline
\hline
             & $\lambda_0$ & $\sigma$ & flux & $V_s$  &     & $\lambda_0$ & $\sigma$ & flux & $V_s$ \\
\hline
\multicolumn{5}{c|}{double-peaked profile in [O~{\sc iii}]$\lambda5007$\AA} &
\multicolumn{5}{c}{double-peaked profile in [O~{\sc iii}]$\lambda4363$\AA} \\
red-shifted  & 5016.17$\pm$0.14 & 3.61$\pm$0.21 & 572$\pm$78 & 0
& red-shifted  & 4363.09$\pm$2.63 & *3.31$\pm$2.31* & *24$\pm$21* & *-568$\pm$181* \\
blue-shifted & 5007.50$\pm$0.35 & 7.62$\pm$0.22 & 2102$\pm$112 & -519$\pm$21
& blue-shifted & 4370.46$\pm$1.98 & *2.79$\pm$1.26* & *24$\pm$18* & *-61$\pm$136* \\

\multicolumn{5}{c|}{broad emission line} &
\multicolumn{5}{c}{double-peaked profile in [Ne~{\sc iii}]$\lambda3970$\AA} \\
H$\alpha$ & 6550.13$\pm$1.71 & 31.25$\pm$0.73 & 2226$\pm$224 & -1137$\pm$78
& red-shifted  & 3874.81$\pm$0.54 & 2.68$\pm$0.45 & *82$\pm$35* & *-90$\pm$42* \\
 &  &  &  & 
& blue-shifted & 3867.58$\pm$1.47 & *4.55$\pm$1.01* & *129$\pm$39* & -651$\pm$114 \\

\multicolumn{5}{c|}{core components in narrow emission lines} &
\multicolumn{5}{c}{extended components in narrow emission lines} \\
H$\alpha$ & 6574.28$\pm$0.14 & 4.69$\pm$0.23 & 716$\pm$74 & *-33$\pm$7*
& H$\alpha$ & 6562.86$\pm$0.82 & 12.70$\pm$0.49 & 967$\pm$98 & -555$\pm$37 \\
H$\beta$  & 4869.84$\pm$0.11 & 3.48$\pm$0.17 & 157$\pm$15 & *-33$\pm$7*
& H$\beta$ & 4861.38$\pm$0.61 & 9.41$\pm$0.36 & 330$\pm$20 & -555$\pm$37 \\
H$\gamma$ & 4348.35$\pm$0.82 & *4.93$\pm$1.23* & *95$\pm$113* & *-14$\pm$57*
& H$\gamma$ & 4339.70$\pm$7.93 & *7.90$\pm$3.48* & *123$\pm$119* & *-621$\pm$548* \\
H$\delta$ & 4104.36$\pm$0.78 & *4.65$\pm$1.16* & *52$\pm$33* & -368$\pm$57
& H$\delta$ & 4096.20$\pm$7.49 & *7.46$\pm$3.29* & *27$\pm$35* & *-964$\pm$548* \\
H$\epsilon$ & 3973.33$\pm$0.80 & *3.60$\pm$1.55* & *55$\pm$78* & *-231$\pm$61*
& H$\epsilon$ & 3966.93$\pm$7.80 & *6.18$\pm$3.14* & *65$\pm$82* & *-715$\pm$589* \\
{[N~{\sc ii}]}$\lambda$6584 & 6593.11$\pm$0.14 & 4.99$\pm$0.16 & 834$\pm$52 & -118$\pm$6
& {[N~{\sc ii}]}$\lambda$6584 & 6580.99$\pm$1.49 & 15.27$\pm$0.83 & 1571$\pm$229 & -671$\pm$68 \\
{[S~{\sc ii}]}$\lambda$6716 & 6726.37$\pm$0.12 & 4.98$\pm$0.10 & 453$\pm$14 & -114$\pm$6
& {[S~{\sc ii}]}$\lambda$6716 & 6720.99$\pm$1.65 & 22.39$\pm$1.22 & 498$\pm$35 & *-354$\pm$74* \\
{[S~{\sc ii}]}$\lambda$6731 & 6740.76$\pm$0.12 & 4.99$\pm$0.10 & 349$\pm$13 & -114$\pm$6
& {[S~{\sc ii}]}$\lambda$6731 & 6735.38$\pm$1.65 & 22.44$\pm$1.23 & 384$\pm$35 & *-354$\pm$74* \\
{[S~{\sc ii}]}$\lambda$4070 & 4073.67$\pm$0.41 & 6.76$\pm$0.51 & 154$\pm$16 & -661$\pm$30
& {[S~{\sc ii}]}$\lambda$4070 & 4064.54$\pm$5.14 & 34.13$\pm$5.30 & *201$\pm$41* & *-1333$\pm$379* \\
{[O~{\sc i}]}$\lambda$6300 & 6308.95$\pm$0.17 & 6.61$\pm$0.20 & 364$\pm$15 & -147$\pm$8
& {[O~{\sc i}]}$\lambda$6300 & 6293.58$\pm$1.19 & 21.74$\pm$0.69 & 429$\pm$6 & -878$\pm$57 \\
{[O~{\sc i}]}$\lambda$6363 & 6372.51$\pm$0.17 & 6.68$\pm$0.20 & 107$\pm$6 & -147$\pm$8
& {[O~{\sc i}]}$\lambda$6363 & 6356.98$\pm$1.21 & 21.96$\pm$0.70 & 126$\pm$6 & -878$\pm$57 \\
{[O~{\sc ii}]}$\lambda$3727 & 3733.17$\pm$0.07 & 2.88$\pm$0.09 & 552$\pm$28 & -98$\pm$6
& {[O~{\sc ii}]}$\lambda$3727 & 3728.18$\pm$0.40 & 6.83$\pm$0.20 & 485$\pm$29 & -499$\pm$32 \\
\multicolumn{5}{c|}{} 
& {[O~{\sc iii}]}$\lambda$5007 & 4991.73$\pm$0.51 & 20.04$\pm$0.32 & 2173$\pm$56 & -1464$\pm$31 \\
 &  &  &  & 
& {[O~{\sc iii}]}$\lambda$4959 & 4943.94$\pm$0.51 & 19.85$\pm$0.32 & 724$\pm$19 & -1464$\pm$31 \\
 &  &  &  & 
& {[Ne~{\sc iii}]} & 3873.33$\pm$1.28 & 19.47$\pm$1.52 & 267$\pm$20 & *-205$\pm$100* \\
\hline
\end{tabular}
\label{data}

{\bf Note:} The first and sixth columns show the emission line labels.
The second and seventh columns show the rest of the central wavelength in units of \AA.
The third and eighth columns show the line width (second moment) in units of \AA.
The fourth and ninth columns show the line flux in units of $10^{-17}{\rm erg/s/cm^2}$. 
The fifth and tenth columns show the shifted velocity of each emission component relative to the red-shifted component of [O~{\sc iii}]$\lambda5007$\AA\ in units of $km/s$.
The ** symbols indicate that the measurements are considered unreliable due to the line widths, fluxes, or velocity shifts being less than five times their corresponding uncertainties. 
\end{table*}

After masking out the emission lines, the observed SDSS spectrum can be well described by the strengthened, broadened, and shifted SSPs (the broadened velocity as the stellar velocity dispersion and the shifted velocity as the velocity to correct for the redshift) plus the power-law component through the Levenberg-Marquardt least-squares minimization technique (the MPFIT package), leading to $\chi_1^2 = SSR_1/Dof_1\sim1.33$, where $SSR_1 = 2488.85$ and $Dof_1 = 1867$ as the summed squared residuals and the degrees of freedom, respectively. 
Here, when the emission lines are being masked out, besides the H$\alpha$ and the [O~{\sc iii}]$\lambda\lambda4959,5007$\AA\ doublet, the other narrow emission lines with rest wavelengths between 3700\AA\ and 7000\AA\ are masked out with full widths at a zero intensity of about ${\rm 1200km/s}$, due to the broad extended components in the narrow emission lines, as shown in the following results. 
Meanwhile, due to the very broad components in the H$\alpha$ and the [O~{\sc iii}]$\lambda\lambda4959,5007$\AA\ doublet, as shown in the following results, the spectrum has been masked within the wavelength ranges from 4900\AA\ to 5050\AA\ and from 6400\AA\ to 6700\AA. 
The best descriptions of the host galaxy contributions and the corresponding emission line spectra are shown in Fig.~\ref{spec}.

Besides the primary model (Model\_1) incorporating the 39 SSPs and a power-law component, a secondary model (Model\_2) is applied considering only the 39 SSPs without considering the AGN continuum emissions. 
Similar as the considerations in \citet{2022MNRAS.511..214N}, through the Levenberg-Marquardt least-squares minimization technique, the Model\_2 applied to determine host galaxy contributions can lead to $\chi_2^2 = SSR_2/Dof_2 = 3765.65/1869\sim2.01$. 
Here, due to the poor fit descriptions of the SDSS spectrum of \obj\ by the Model\_2 function, the best descriptions are not presented in this manuscript. 
However, similar as what we have recently done in \citet{2024ApJ...961...82Z}, through the $SSR$ and $Dof$ related to the Model\_1 and Model\_2 functions of the F-test technique, the calculated value $F_p$ is about 
\begin{equation}
   F_p = \frac{(SSR_2 - SSR_1)/(Dof_2 - Dof_1)}{SSR_1/Dof_1} \sim 478.89
\end{equation}.
Meanwhile, considering $Dof_2 - Dof_1 = 2$ and $Dof_1 = 1867$ as the number of $Dof$ of the $F$ distribution numerator and denominator, the expected value is about $F_e\sim14.50$ from the statistical F-test with a confidence level of about $5\sigma$.
The $F_p\sim478.89$ significantly higher than $F_e\sim14.50$ indicates that the model\_1 with considerations of the AGN power-law continuum emissions should be preferred with a confidence level much higher than $5\sigma$.

\begin{figure}
\centering\includegraphics[width = 9cm,height=5cm]{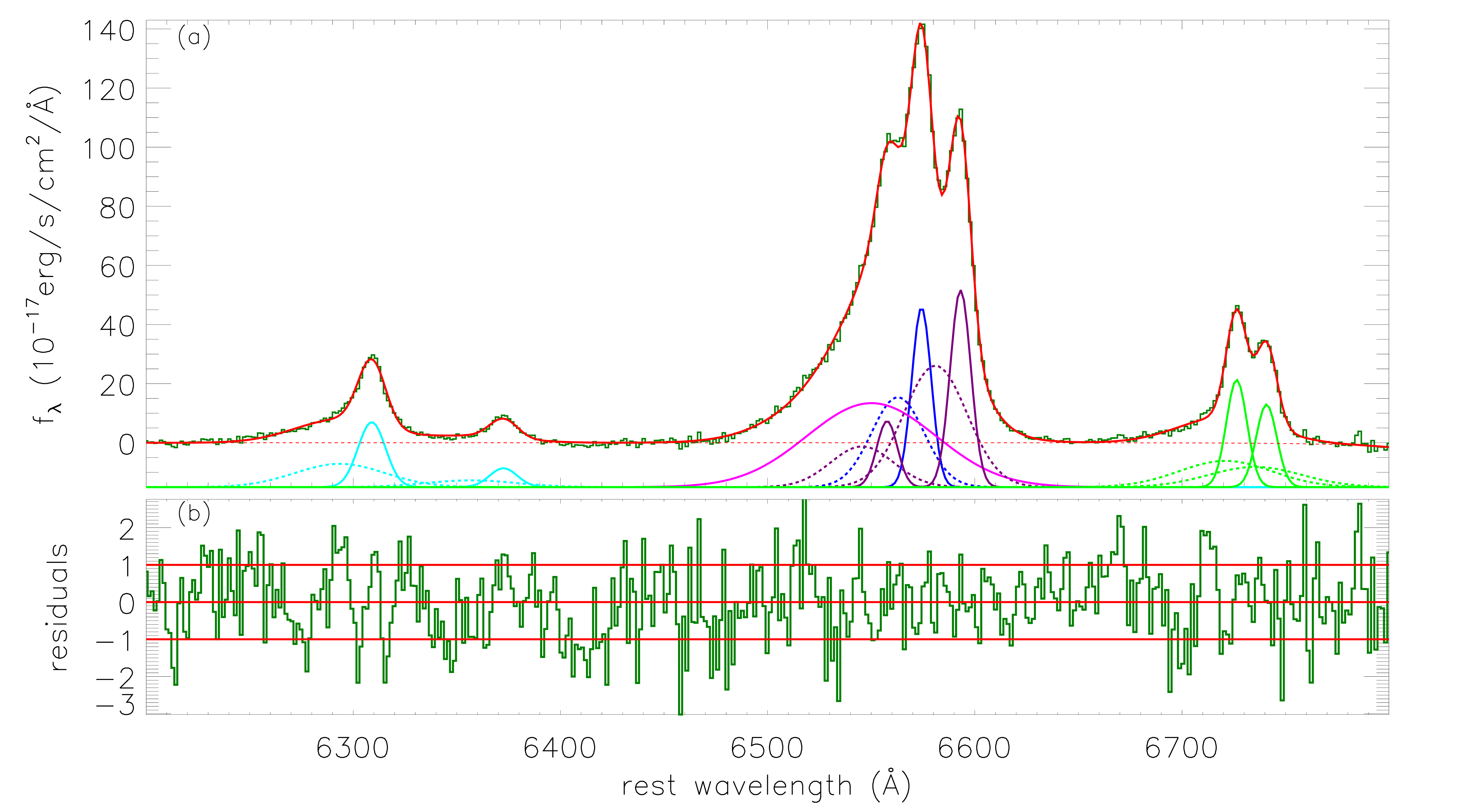}
\centering\includegraphics[width = 9cm,height=5cm]{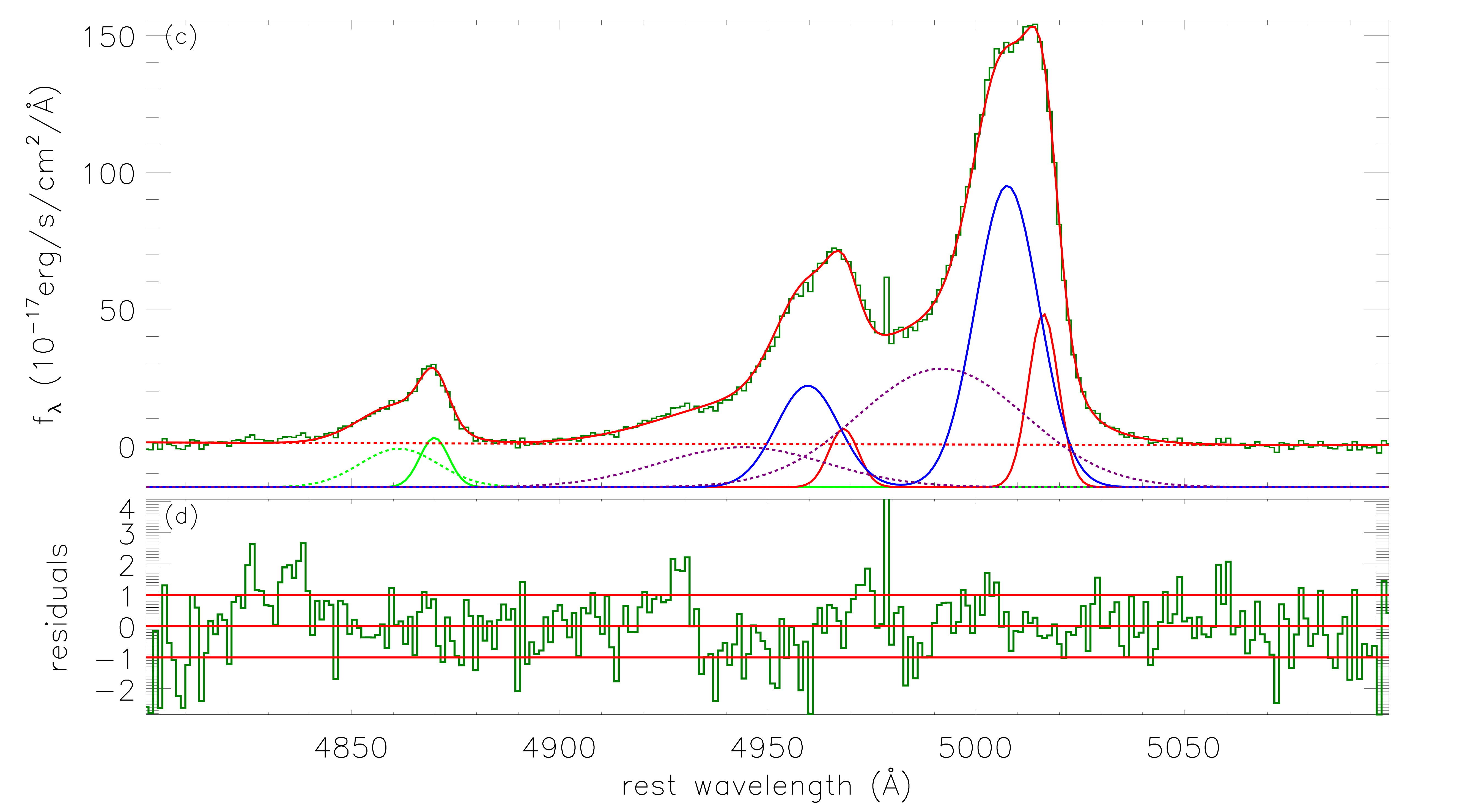}
\centering\includegraphics[width = 9cm,height=5cm]{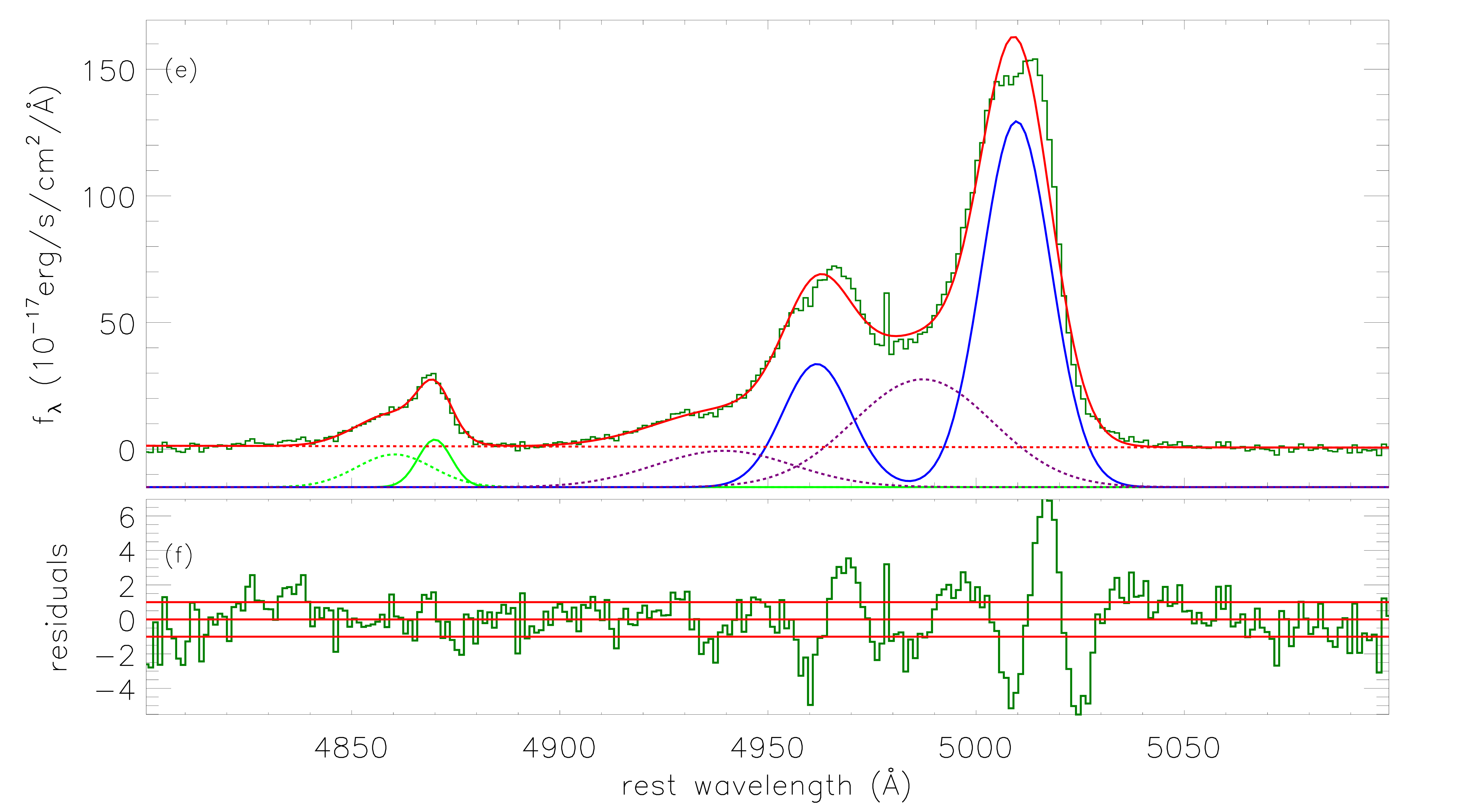}
\caption{The best-fitting results and the corresponding residuals to the emission lines around H$\alpha$ (panels (a) and (b)), and around H$\beta$ (panels (c) and (d)), and the bad-fitting results and the corresponding residuals to the emission lines around H$\beta$ (panels (e) and (f)). 
In panels (a), (c), and (e), the solid dark green lines show the line spectrum, and the solid red lines show the best-fitting results. 
The horizontal dashed red line shows $f_\lambda = 0$. 
In panel (a), the solid and dashed blue lines show the determined core and extended components in narrow H$\alpha$, 
and the solid magenta line shows the determined component in broad H$\alpha$. 
The solid and dashed purple lines show the determined core and extended components in the [N~{\sc ii}] doublet. 
The solid and dashed cyan lines show the determined core and extended components in the [O~{\sc i}] doublet. 
The solid and dashed green lines show the determined core and extended components in the [S~{\sc ii}] doublet. 
In panel (c), the solid and dashed green lines show the determined core and extended components in narrow H$\beta$. 
The solid blue and red lines show the determined blue-shifted and red-shifted emission profiles in the [O~{\sc iii}] doublet. 
The dashed purple lines show the determined extended components in the [O~{\sc iii}] doublet. 
In panel (e), the solid blue line shows the determined core components in the [O~{\sc iii}] doublet. 
The other line styles have the same meanings as those in panel (c).
In panels (b), (d), and (f), the corresponding residuals are shown in dark green, with the horizontal red lines showing residuals $= 0,\pm1$.}
\label{line}
\end{figure}

\begin{figure}
\centering\includegraphics[width = 9cm,height=5cm]{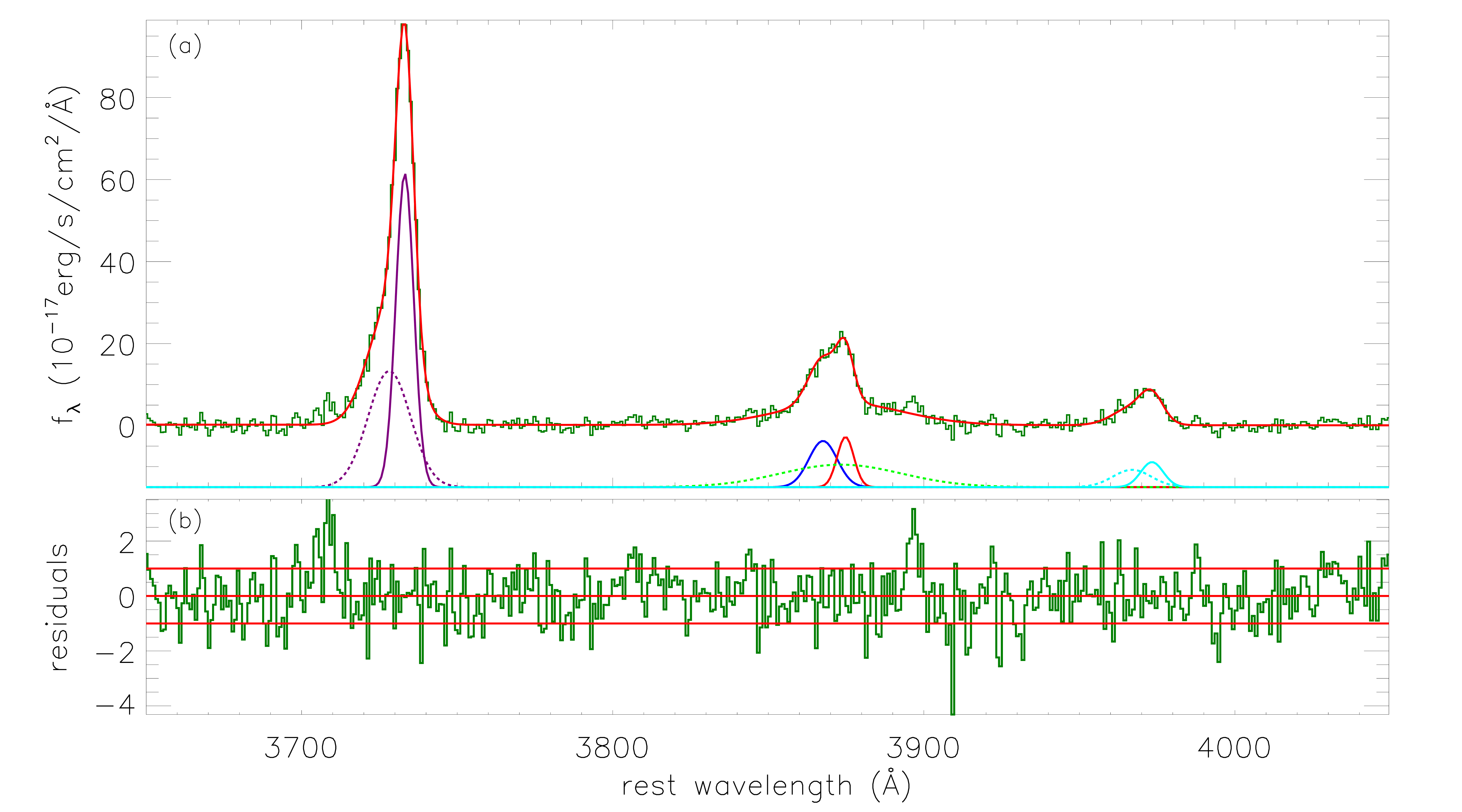}
\centering\includegraphics[width = 9cm,height=5cm]{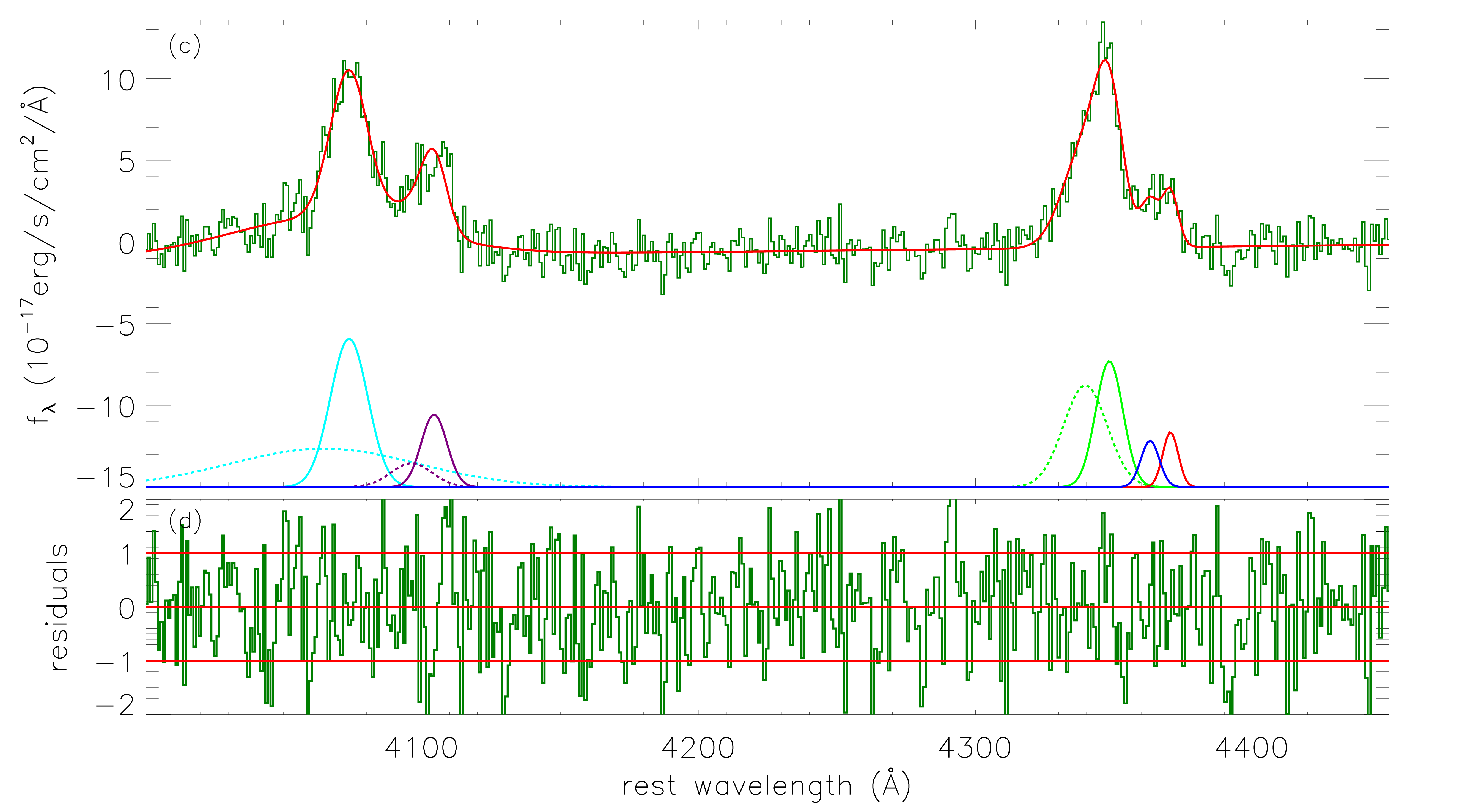}
\caption{The best-fitting results and the corresponding residuals to the emission lines around [O~{\sc ii}] (panels (a) and (b)) and around H$\gamma$ (panels (c) and (d)). 
In panels (a) and (c), the solid dark green lines show the line spectrum, and the solid red lines show the best-fitting results. 
In panel (a), the solid and dashed purple lines show the determined core and extended components in [O~{\sc ii}]$\lambda3727$\AA. 
The solid and dashed cyan lines show the determined core and extended components in H$\epsilon$. 
The solid blue and red lines show the determined blue-shifted and red-shifted components in [Ne~{\sc iii}]. 
The dashed green line shows the determined extended component in [Ne~{\sc iii}]. 
In panel (c), the solid and dashed green lines show the determined core and extended components in H$\gamma$. 
The solid and dashed purple lines show the determined core and extended components in narrow H$\delta$. 
The solid blue and red lines show the determined blue-shifted and red-shifted emission profiles  in [O~{\sc iii}]$\lambda4363$\AA. 
The solid and dashed cyan lines show the determined core and extended components in [S~{\sc ii}]$\lambda4070$\AA. 
In panels (b) and (d), the corresponding residuals are shown, with the horizontal red lines showing residuals $ = 0,\pm1$.}
\label{o2}
\end{figure}

\begin{figure}
\centering
\includegraphics[width = 9cm,height=5cm]{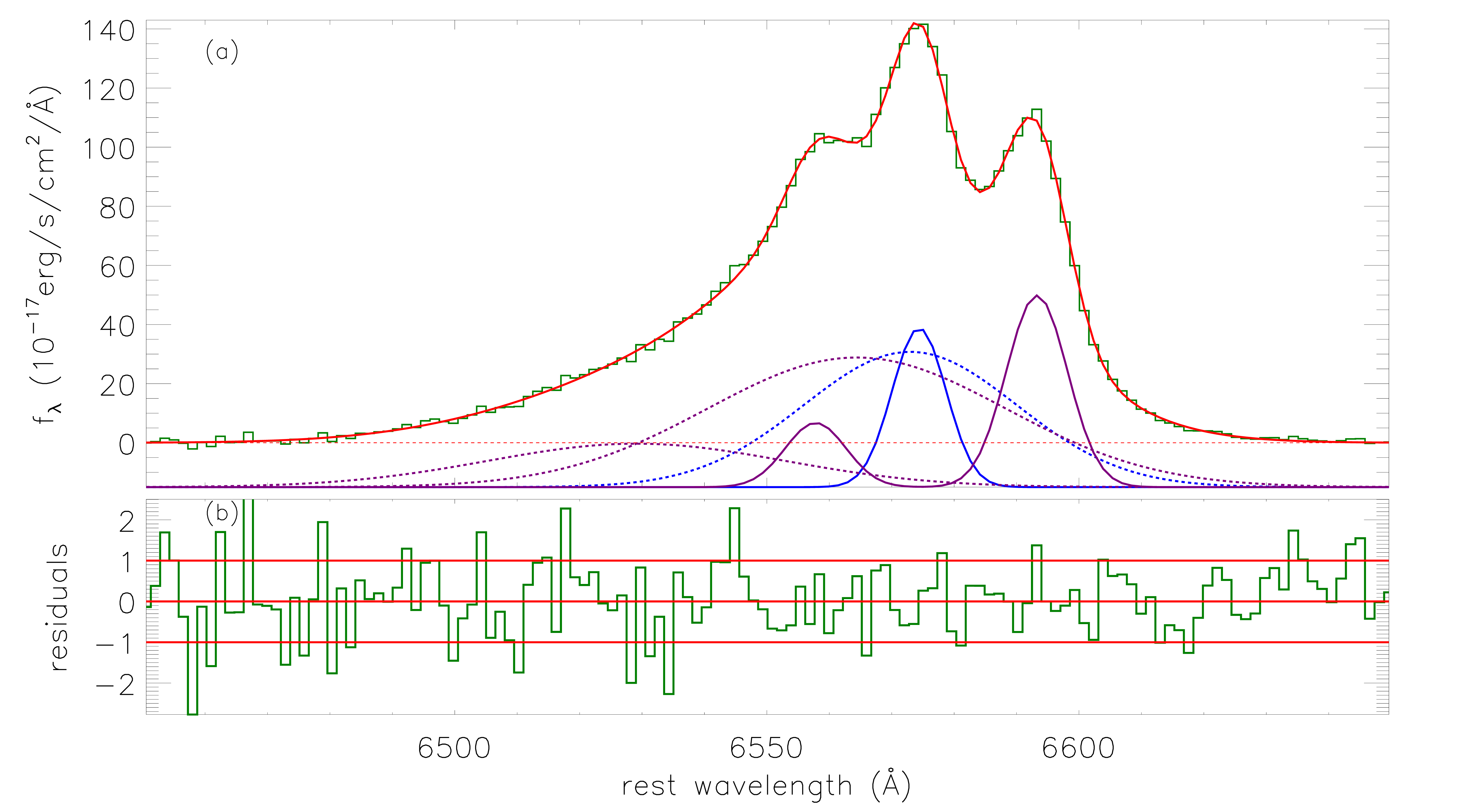}
\includegraphics[width = 9cm,height=5cm]{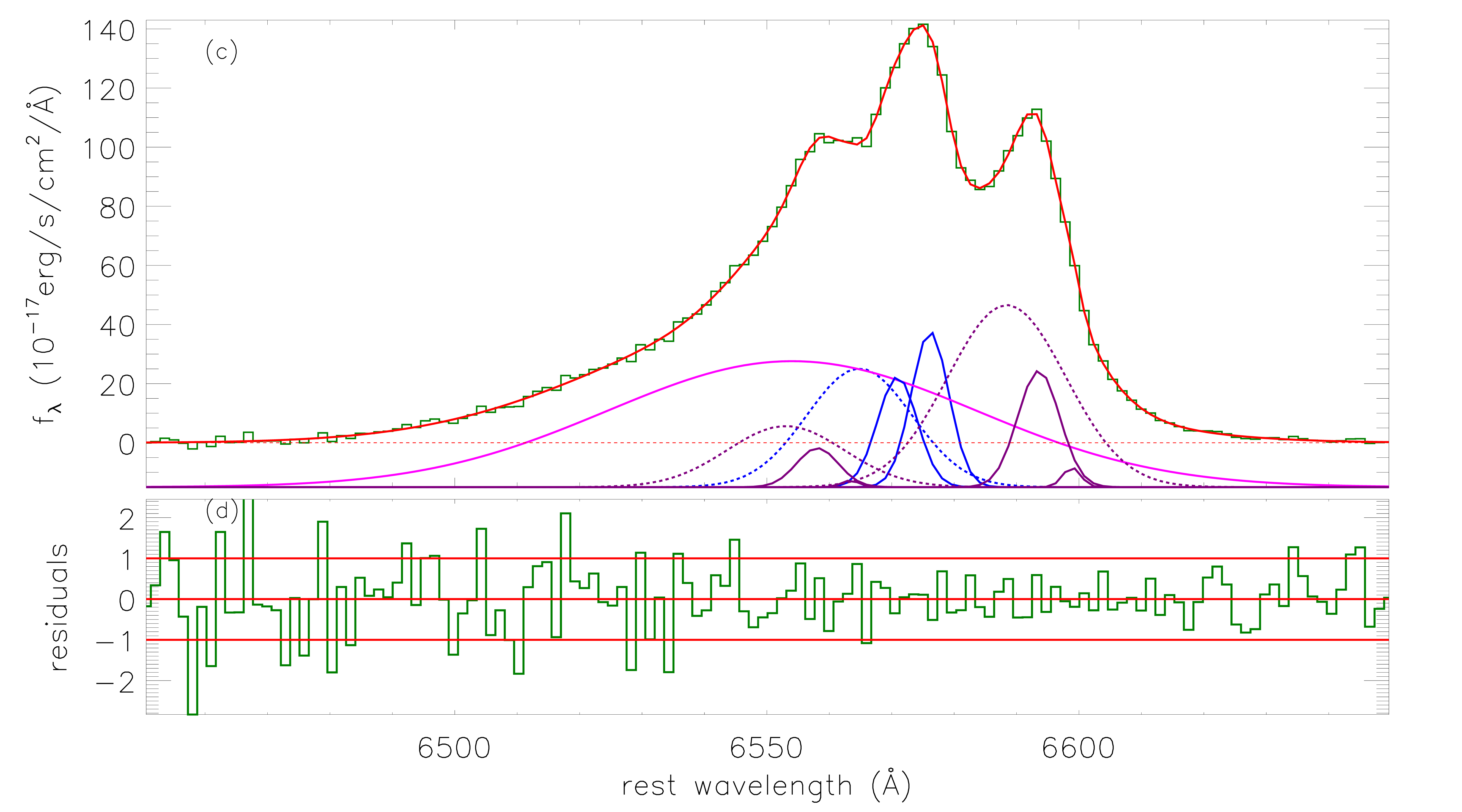}
\includegraphics[width = 9cm,height=5cm]{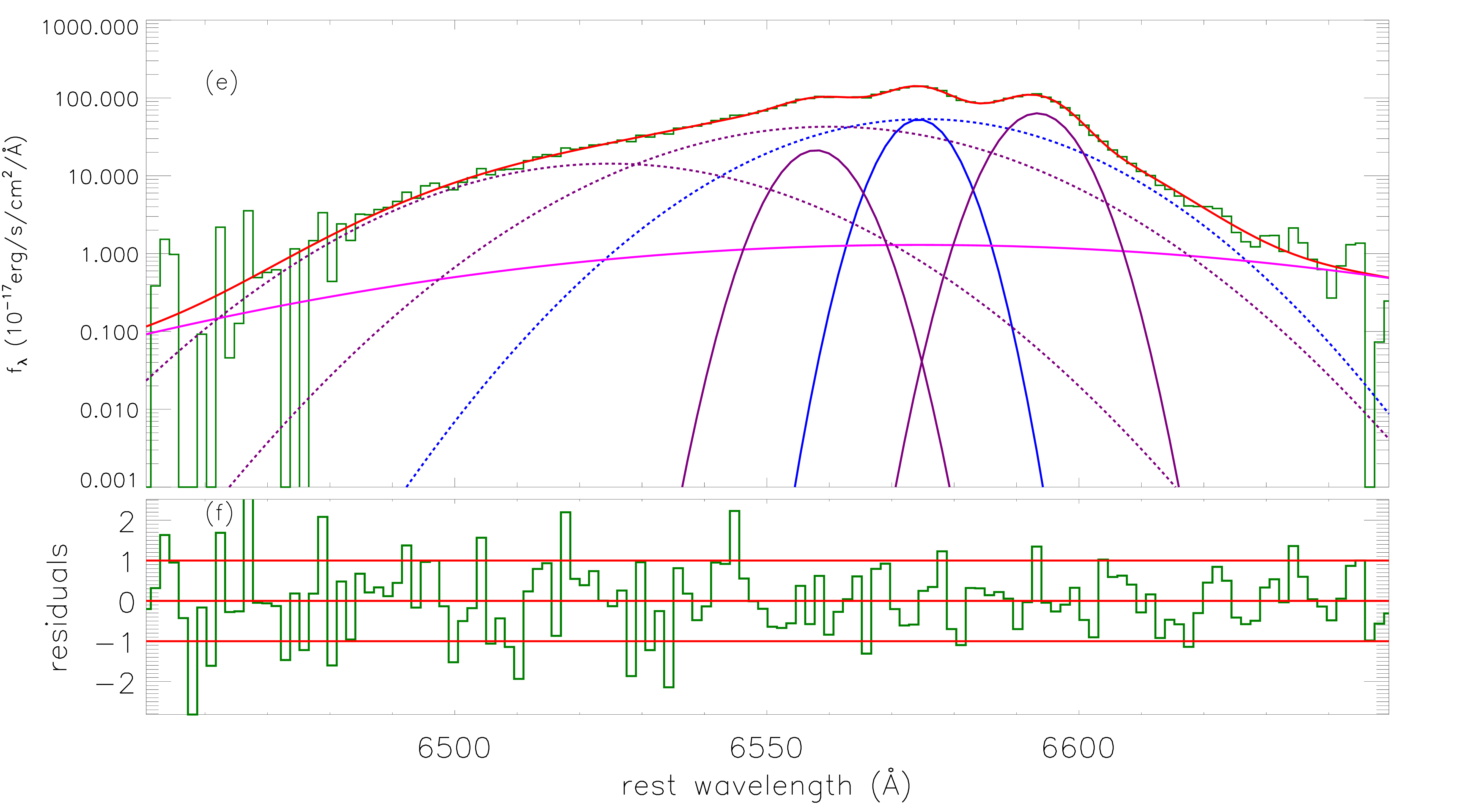}
\caption{The best-fitting results and the corresponding residuals to H$\alpha$ and [N~{\sc ii}] emission lines for Model\_a (panels (a) and (b)), Model\_c (panels (c) and (d)), and Model\_d (panels (e) with the y-axis shown on a logarithmic scale and (f)).
All line styles have the same meanings as those in panels (a) and (b) in Fig.~\ref{line}.}
\label{fig:Ha_fix}
\end{figure}

\begin{table}
\footnotesize
\caption{Line parameters of the H$\alpha$ and {[N~{\sc ii}]}$\lambda$6584 emission lines in \obj\ for Model\_a, Model\_c, and Model\_d}
\centering
\begin{tabular}{lccc}
\hline
\hline
             & $\lambda_0$ & $\sigma$ & flux \\
\hline
\multicolumn{4}{c}{Model\_a : $SSR_{a} = 111.46, Dof_{a} = 121$} \\
core H$\alpha$     & 6574.39$\pm$0.17 & 4.33$\pm$0.28 & 584$\pm$68 \\
extended H$\alpha$ & 6572.97$\pm$1.02 & 17.07$\pm$0.87 & 1958$\pm$291 \\
core {[N~{\sc ii}]}$\lambda$6584  & 6593.26$\pm$0.12 & 4.90$\pm$0.17 & 796$\pm$52 \\
extended {[N~{\sc ii}]}$\lambda$6584 & 6564.39$\pm$1.46 & 23.87$\pm$0.55 & 2623$\pm$184 \\
\hline
\multicolumn{4}{c}{Model\_c : $SSR_{c} = 89.51, Dof_{c} = 112$} \\
core H$\alpha$     & 6576.15$\pm$2.06 & 3.00$\pm$0.82  & *395$\pm$375* \\
core H$\alpha$     & 6570.91$\pm$3.30 & *3.18$\pm$1.66*  & *296$\pm$458* \\
extended H$\alpha$   & 6564.82$\pm$5.10 & *8.42$\pm$3.15*  & *846$\pm$509* \\
broad H$\alpha$      & 6554.03$\pm$0.75 & 29.55$\pm$0.36 & 3156$\pm$120 \\
core {[N~{\sc ii}]}$\lambda$6584  & 6598.84$\pm$0.72 & *1.42$\pm$0.85* & *24$\pm$31* \\
core {[N~{\sc ii}]}$\lambda$6584  & 6593.55$\pm$0.38 & 3.36$\pm$0.67 & *332$\pm$121* \\
extended {[N~{\sc ii}]}$\lambda$6584 & 6588.44$\pm$0.75 & 9.43$\pm$0.61 & 1456$\pm$196 \\
\hline
\multicolumn{4}{c}{Model\_d : $SSR_{d} = 105.49, Dof_{d} = 119$} \\
core H$\alpha$     & 6574.35$\pm$0.16 & 4.26$\pm$0.29  & 565$\pm$69 \\
extended H$\alpha$ & 6575.31$\pm$1.62 & 17.80$\pm$0.58 & 2385$\pm$308 \\
broad H$\alpha$    & 6574.35 & *53.86$\pm$29.61* & *175$\pm$69* \\
core {[N~{\sc ii}]}$\lambda$6584  & 6593.31$\pm$0.13 & 4.82$\pm$0.15  & 769$\pm$46 \\
extended {[N~{\sc ii}]}$\lambda$6584 & 6560.09$\pm$3.11 & 20.84$\pm$2.06 & 2230$\pm$237 \\
\hline
\end{tabular}
\label{model}
{\bf Note:} {The content and units in each column are the same as those in Table~\ref{data}, except for the absence of $V_s$.
The ** symbols indicate that the measurements are considered unreliable due to the line widths or fluxes being less than three times their corresponding uncertainties. 
}
\end{table}

For the second step, after the host galaxy contributions are well determined, the emission lines measured from the spectrum (shown in the bottom panel of Fig.~\ref{spec}), derived by subtracting the combined host galaxy contributions and AGN continuum emissions from the SDSS spectrum, can be described by multiple Gaussian components.
Here, the following functions are applied to measure the emission lines in the line spectrum of \obj.

The emission lines within the rest wavelength range from 4800\AA\ to 5100\AA\ and from 6200\AA\ to 6800\AA, including H$\beta$, [O~{\sc iii}] doublet, [O~{\sc i}] doublet, H$\alpha$, [N~{\sc ii}] doublet, and [S~{\sc ii}] doublet, are primarily considered and measured. 
Each narrow emission line is fitted with two Gaussian functions, representing a core component and a shifted extended component, except for the [O~{\sc iii}] doublet, which displays distinct double-peaked profiles and requires three Gaussian functions: two for the double peaks and one for the extended component. 
For H$\alpha$ and H$\beta$, in addition to the narrow emission components (both the core and the extended components), one broad Gaussian function is applied to each to describe its broad component.

When the above-mentioned model functions are applied, the following restrictions are adopted. 
First, the corresponding components within each doublet (including H$\alpha$ and H$\beta$) share the same redshift and line width in velocity space. 
Second, the theoretical value of 3:1 has been set to the flux ratio of the [O~{\sc iii}] and [N~{\sc ii}] doublets' corresponding components. 
Third, all Gaussian components are constrained to have line intensities not smaller than zero.

Then, based on the model functions above, through the Levenberg-Marquardt least-squares minimization technique, the emission lines within the rest wavelength range from 4800\AA\ to 5100\AA\ and from 6200\AA\ to 6800\AA\ can be described (Model\_G). 
The corresponding fitting results and residuals (line spectrum minus the best-fitting results, and then divided by the uncertainties of the line spectrum) are shown in panels (a), (b), (c), and (d) in Fig.~\ref{line}, with the determined ${\chi_{G}}^2 = SSR_{G}/Dof_{G} = 715.26/621\sim1.15$.
The measured line parameters and the corresponding uncertainties are listed in Table~\ref{data}. 
According to the fitting results, the Gaussian flux of the broad H$\beta$ component is smaller than the corresponding uncertainty, suggesting that there is no significant broad H$\beta$ emission in the spectrum.
Here, it is worthy to note that the redshift for \obj\ provided by the SDSS pipeline is 0.120474. 
However, considering the SSP method determined a shifted velocity of $+399.96\pm14.57$km/s through the host galaxy absorption features, the corrected redshift should be $0.121807\pm4.86\times10^{-5}$. 
Therefore, the listed central wavelength of each emission component in Table~\ref{data} is the value after accepting $z=0.121807$.

Meanwhile, besides the emission lines around H$\alpha$ and H$\beta$, the other emission lines are measured by similar model functions as described above. 
Their best-fitting results and the corresponding residuals are shown in Fig.~\ref{o2}, and the measured line parameters are also listed in Table~\ref{data}.

Before proceeding further, three points should be noted. 
First, besides the three Gaussian functions applied, only two Gaussian functions have also been considered (Model\_O) to describe the [O~{\sc iii}]$\lambda5007$\AA\ ([O~{\sc iii}]$\lambda4959$\AA). 
As shown in panels (e) and (f) in Fig.~\ref{line}, the larger residuals around the [O~{\sc iii}] doublet, along with the corresponding ${\chi_{O}}^2= SSR_{O}/Dof_{O} = 1244.03/624\sim1.99$, are significantly worse than those of the Model\_G.
Based on the $SSR$s and $Dof$s values for Model\_G and Model\_O, the expected value is about $F_e\sim10.88$ from the statistical F-test with a confidence level of about $5\sigma$.
The $F_p\sim153.03$ significantly higher than $F_e\sim10.88$ indicates that the model\_G with considerations of three Gaussian functions to describe each of the [O~{\sc iii}] doublet emission lines should be preferred with a confidence level much higher than $5\sigma$.

Second, in order to focus more closely on the H$\alpha$ and [N~{\sc ii}] doublet, a model using a narrow component plus an extended component (Model\_a) is initially considered to describe each of the H$\alpha$ and [N~{\sc ii}] emission lines within the rest wavelength range from 6450\AA\ to 6650\AA.
The best-fitting results and the corresponding residuals of the emission lines are shown in panels (a) and (b) of Fig.~\ref{fig:Ha_fix}. 
The measured line parameters and the corresponding uncertainties are listed in Table~\ref{model}, with ${\chi_{a}}^2= SSR_{a}/Dof_{a} = 111.46/121\sim0.92$.
Then, in order to explore the effect of adding a broad H$\alpha$ to Model\_a to form Model\_b, the $SSR$ and $Dof$ are recalculated within the same rest wavelength range, leading to ${\chi_{b}}^2= SSR_{b}/Dof_{b} = 98.33/118\sim0.83$, while the best-fitting results, the corresponding residuals, and measured line parameters with their corresponding uncertainties remain those already shown in panels (a) and (b) of Fig.~\ref{line} and listed in Table~\ref{data}. 
By applying the F-test technique to the $SSR$s and $Dof$s of Model\_a and Model\_b, the calculated $F_p\sim5.25$, which corresponds to a confidence level of 99.81$\%$ (3.11$\sigma$), indicating that Model\_b with considerations of a broad H$\alpha$ component is strongly preferred over Model\_a.

\begin{figure*}
\centering\includegraphics[width = 18cm,height=12cm]{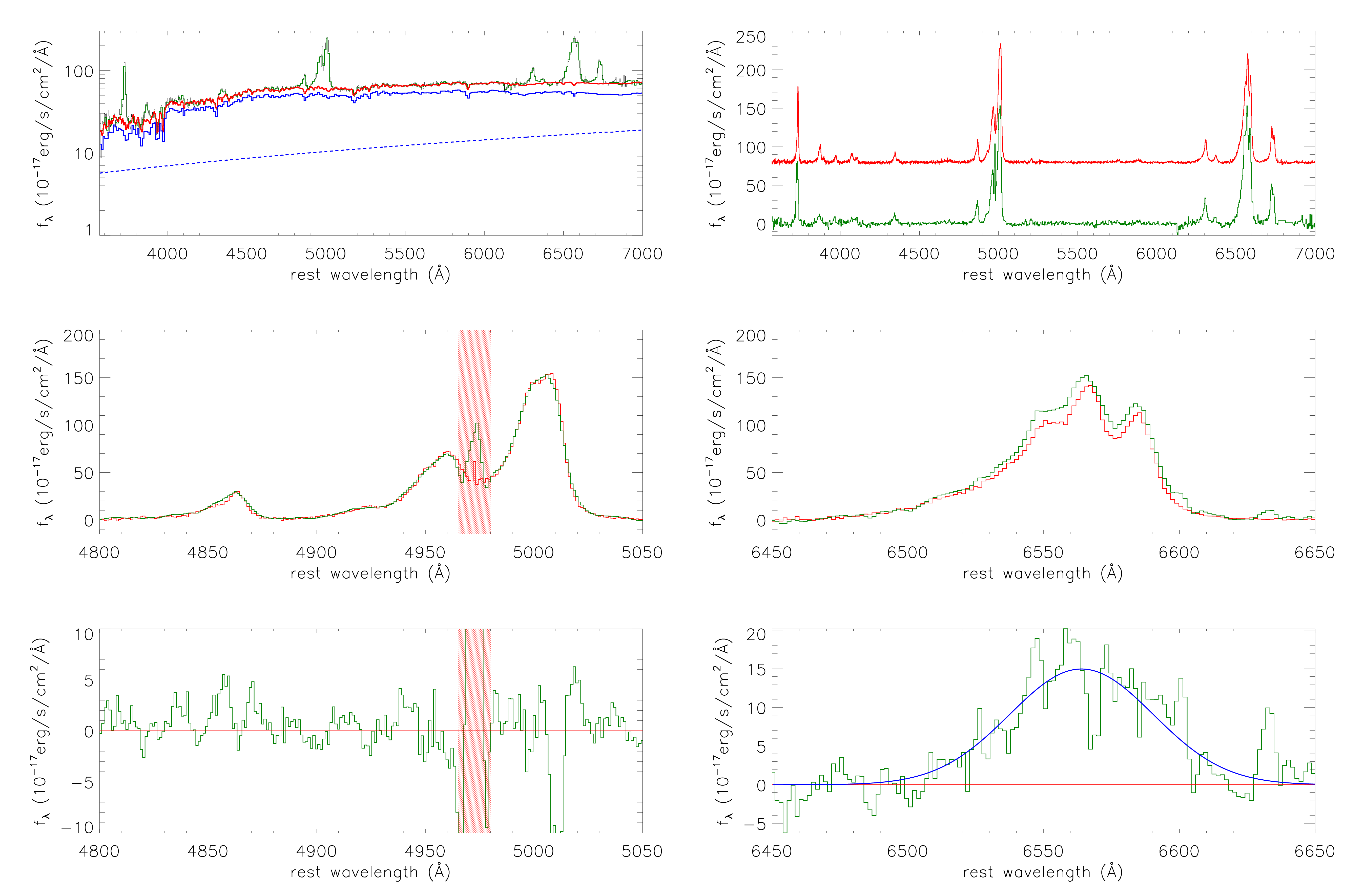}
\caption{Comparative analysis of the emission-line spectra of \obj\ between the SDSS and SP2 observations. 
The top left panel shows the best descriptions of the host galaxy contributions for the SP2 spectrum. 
The line styles have the same meanings as those in the top panel of Fig.~\ref{spec}.
The top right panel shows the comparison of the emission-line spectra after subtracting the host galaxy contributions and AGN continuum emissions between the SP2 (dark green) and SDSS (red; shifted upward by 80 along the y-axis for a clearer comparison).
The middle panels show the detailed emission-line spectra within the top right panel between the SP2 (dark green) and SDSS (red) around H$\beta$ and H$\alpha$, respectively.
The bottom panels show the spectral differences (SP2 - SDSS) around H$\beta$ and H$\alpha$, respectively.
The horizontal solid red line shows $f_\lambda=0$. 
The red semi-transparent regions in the middle left and bottom left panels highlight the significant differences between the SP2 and SDSS spectra, which may be attributed to poor observational quality. 
The solid blue line in the bottom right panel indicates that a single Gaussian function provides a good description of the difference between the SP2 and SDSS spectra.}
\label{dif}
\end{figure*}

Moreover, a direct spectral comparison is also performed to independently confirm the existence of broad H$\alpha$.
In order to further confirm the broad component (not the extended shifted component) in broad H$\alpha$ but no broad component in broad H$\beta$, the variability properties of the emission lines are carefully checked between the SDSS spectrum and the reported spectrum (hereafter, SP2) in \citet{2012RMxAA..48....9T}. 
The SSP method is applied again to determine the host galaxy contributions in the SP2 spectrum after considering the similar 39 SSPs and AGN continuum emissions, as we have done for the SDSS spectrum of \obj. 
The corresponding best-fitting results are shown in the top left panel in Fig.~\ref{dif}. 
Then, after the subtractions of the host galaxy and the AGN continuum contributions, emission line residuals from both spectra are shown in the top right panel of Fig.~\ref{dif}, with detailed emission-line spectra around the H$\beta$ and H$\alpha$ regions shown in the middle panels.
The bottom panels of Fig.~\ref{dif} show the corresponding difference spectra (the line spectrum from the SP2 spectrum minus the one from the SDSS spectrum) around the H$\beta$ and H$\alpha$ regions.

Here, in order to correct the effects of aperture sizes, a scale factor of 1.286 (= 8106.4 / 6302.9) has been applied to the SDSS line spectrum, due to the total line flux (in units of $10^{-17}{\rm erg/s/cm^2}$) of the [O~{\sc iii}] doublet at about 8106.4 and 6302.9 in the SP2 spectrum and in the SDSS spectrum, respectively. 
In the middle and bottom panels of Fig.~\ref{dif}, it is clear that there are no variability components in narrow emissions, but there is an apparent variability component around H$\alpha$. 
Such a variability component is characteristic of BLR emissions rather than NLR ones. 
If the variability component were instead attributed to the extended [N~{\sc ii}] doublet, a comparable feature should also appear near the [O~{\sc iii}] doublet, which is not observed. 
These pieces of evidence do not support a pure origin from narrow and extended components.
Therefore, there is a reliable broad H$\alpha$ emission component from the central BLRs in \obj.

Third, in order to investigate whether the H$\alpha$ and [N~{\sc ii}] emission lines exhibit double-peaked profiles, on the basis of Model\_b, an additional narrow component is  added to each of the narrow H$\alpha$ and [N~{\sc ii}] lines (Model\_c) within the rest wavelength range from 6450\AA\ to 6650\AA.
The best-fitting results and the corresponding residuals of the emission lines are shown in panels (c) and (d) of Fig.~\ref{fig:Ha_fix}. 
The measured line parameters and the corresponding uncertainties are listed in Table~\ref{model}, with ${\chi_{c}}^2= SSR_{c}/Dof_{c} = 89.51/112\sim0.80$.
Although Model\_c provides a statistically acceptable ${\chi_{c}}^2$, the majority of derived second moments and fluxes of Gaussian functions are unreliable, making the model unnecessary to be further tested by the F-test and therefore discarded.
This indicates that adding the additional narrow components does not provide a preferred fit and further supports that, among the strong emission lines near H$\alpha$ and H$\beta$, only the [O~{\sc iii}] doublet exhibits clear double-peaked profiles.

Then, based on the measured line parameters of the emission components, the shifted velocity of each emission component relative to the red-shifted component in [O~{\sc iii}]$\lambda5007$\AA\ in \obj\ can be calculated, as listed in Table~\ref{data}. 
The uncertainties of shifted velocities are determined by the uncertainties of the central wavelengths of the emission components, which are determined by the uncertainty of the SSP method. 
As shown in Table~\ref{data}, the broad H$\alpha$ exhibits a significant blueshift relative to the systemic velocity of the host galaxy. 
To rule out the possibility that this blueshift is caused by unconstrained fitting, the central wavelength of the broad H$\alpha$ has been tied to that of the narrow H$\alpha$, within the rest wavelength range from 6450\AA\ to 6650\AA\ (Model\_d).
The best-fitting results and the corresponding residuals of the emission lines are shown in panels (e) and (f) of Fig.~\ref{fig:Ha_fix} with the y-axis of panel (e)  shown on a logarithmic scale to clearly display the broad H$\alpha$ component.
The measured line parameters and the corresponding uncertainties are listed in Table~\ref{model}, with ${\chi_{d}}^2= SSR_{d}/Dof_{d} = 105.49/119\sim0.89$.
The derived flux of the broad H$\alpha$ (= 175, in units of $10^{-17}{\rm erg/s/cm^2}$) is negligible and unreliable, while the second moments of the extended components of H$\alpha$ and [N~{\sc ii}]$\lambda6584$ reach 17.80\AA\ and 20.84\AA, respectively, which are comparable to that of the extended component of [O~{\sc iii}]$\lambda5007$ in Model\_G.
These results indicate that tying the central wavelength of the broad H$\alpha$ to that of the narrow H$\alpha$ does not support the existence of a broad H$\alpha$.
The substantial blueshift of the broad H$\alpha$ may thus be associated with outflows.

\section{Necessary Discussions} \label{discussions}

\subsection{Observational characteristics}

Based on the analyses presented above, we note four key observational characteristics of \obj.
\begin{itemize}
    \item The presence of a prominent double radio jet structure \citep{2003ApJ...584..135L} indicating nuclear activity.
    \item A variable broad H$\alpha$ is detected between two spectra separated by several years, while no broad H$\beta$ is observed, consistent with a typical Type-1.9 AGN classification, as discussed by \citet{1981ApJ...249..462O} for the first time.
    \item The broad H$\alpha$ exhibits a significantly higher relative shifted velocity (-1137 km/s) than those of the narrow emission lines ($\sim$-650 to -500 km/s), as shown in Table~\ref{data}. 
    \item The double-peaked [O~{\sc iii}] profiles are strongly asymmetric, with flux ratios between the blue-shifted and red-shifted components being about 1.71, while other emission lines rarely exhibit double-peaked profiles, as shown in Fig.~\ref{line} and Fig.~\ref{o2}.
\end{itemize}

\subsection{Hypothetical scenarios}

To interpret these characteristics, based on the best-fitting results of the spectrum of \obj\ from the previous sections, particularly focusing on the fundamental properties of the fitted double-peaked [O~{\sc iii}] doublet, 
three possible scenarios are considered to explain the origin of the double-peaked [O~{\sc iii}] doublet, such as dual AGN merger systems, rotating NLRs, and AGN-driven biconical outflows.

\subsubsection{A dual AGN merger system}

Binary supermassive black holes \citep{2010ApJ...715L..30L} or dual AGN \citep{2009ApJ...698..956C, 2009ApJ...705L..76W, 2012ApJS..201...31G, 2015ApJ...813..103M, 2018ApJ...854..169L} may trigger the double-peaked narrow emission line profiles in galaxies. 
If sufficient gas is available, both supermassive black holes could power AGNs during the merger phase, and when the extended NLRs do not overlap, the spectrum will show two sets of AGN emission lines, including the [O~{\sc iii}] doublet \citep{2009ApJ...698..956C}. 
Theoretically, it is predicted that dual supermassive black holes can form during galaxy mergers, but obtaining observational evidence is very difficult, especially since the two black holes may be obscured by dust and gas, making observations challenging \citep{2012ApJ...744....7B}. 
Based on the [O~{\sc iii}] double-peaked profiles, \citet{2025ApJS..277...49Z} conducted a search for dual AGN among 11,557 quasars and identified 62 Type-1 AGN with reliable double-peaked [O~{\sc iii}] lines. 
However, only 7 of them exhibit signatures of mergers and are considered candidates for dual AGN systems on kpc scales. 
\citet{2025MNRAS.539L..24C} analyzed seven kpc-scale dual-core systems with double-peaked narrow emission lines and found that orbital kinematics is unlikely to be the dominant origin, as three systems clearly rule it out and the rest are better explained by alternative mechanisms.
Therefore, at kpc scales, the likelihood that double-peaked narrow emission line profiles originate from orbital kinematics is generally low.
Moreover, dual AGN merger systems should exhibit the following characteristics: (1) All narrow emission lines show double-peaked profiles; 
(2) The red-shifted and blue-shifted components have similar fluxes and line ratios, as discussed in \citet{2009ApJ...705L..20X}.

Although a dual AGN system consisting of a Type-1.9 AGN with a BLR orbiting an AGN without a BLR could, in principle, account for the [O~{\sc iii}] double-peaked profiles of \obj, which can be considered as originating from the two NLRs of a pair of AGNs, several characteristic inconsistencies challenge this interpretation. 
Specifically, the blue-shifted component of the [O~{\sc iii}] doublet is significantly stronger than the red-shifted one, and most other emission lines lack double-peaked profiles. 
In addition, if this object were indeed a dual AGN merger system, the shifted velocity of the broad H$\alpha$ would be expected to coincide with the shifted velocity of one of the narrow components in the double-peaked profiles. 
However, as shown in Table~\ref{data}, such a characteristic is not present in this object. 
Therefore, the scenario of a merging dual AGN is unlikely to be responsible for the origin of the double-peaked profiles in \obj.
However, without further evidence, we cannot completely rule out this possibility. 
Future studies could use the Very Long Baseline Array \citep{2015ApJ...813..103M, 2018ApJ...854..169L} and/or long-slit spectroscopy \citep{2016ApJ...832...67N, 2018ApJ...867...66C} to search for more evidence.

\subsubsection{A rotating NLR}

The double-peaked profiles of emission lines may arise from ionized gas rotating at different velocities within galaxy disks \citep{2012MNRAS.422.1394E, 2019MNRAS.487.3007K}. 
In pure geometrical terms, such systems can be represented by a model consisting of a flat and extended disk component plus a central spherical core \citep{2009ApJ...705L..20X}. 
The blue-shifted and red-shifted narrow emission lines are generally thought to originate from the classical NLR, which is disk-like in structure \citep{2005ApJ...627..721G}, while any extended emission, if present, may arise from a spherical region in the innermost NLR or the outer portion of the BLR. 
The observed line profiles and the relative positions of the blue-shifted and red-shifted peaks are highly sensitive to the orientation of the disk plane and its outer radius \citep{2012ApJ...744....7B}. 
Based on axisymmetric rotating disk models, \citet{2023A&A...670A..46M} found that the peak separation increases with disk inclination. 
Observationally, such double-peaked profiles are characterized by relatively low velocities and velocity dispersions, with typical values smaller than 400 and 500$km/s$, respectively \citep{2023MNRAS.524.2224L}, and exhibit high symmetry, with flux ratios between the blue-shifted and red-shifted components ranging from 0.75 to 1.25 \citep{2012ApJ...752...63S}.

However, for \obj, not all emission lines exhibit a double-peaked profile, in particular, the [O~{\sc iii}] doublet shows a pronounced asymmetry.
The flux ratios between the blue-shifted and red-shifted components of the [O~{\sc iii}] doublet reach as high as 1.71, which exceeds the typical range (0.75-1.25) found in symmetrical double-peaked AGNs in \citet{2012ApJ...752...63S}, indicating a possible deviation from the standard rotating disk scenario.
Therefore, although the possibility that the double-peaked profiles of \obj\ are caused by a rotating NLR cannot be entirely ruled out based on current evidence, the rotating NLR scenario alone is unlikely to fully account for the observed features, and alternative mechanisms may be dominant.

\subsubsection{An AGN-driven biconical outflow}

AGN-driven biconical outflows \citep{2004AJ....127..606W, 2021MNRAS.508.1305Y} can also trigger double-peaked narrow emission line profiles and be collimated by a thick torus, which provides the necessary geometry for producing the characteristic bicone structure \citep{1985ApJ...297..621A}. 
Observations indicate that such outflows are randomly oriented with respect to the stellar disk of their host galaxies \citep{2013ApJS..209....1F}. 
More specifically, the direction of the ionized gas outflows often forms a random angle with the galactic plane \citep{2010MNRAS.402..819S, 2011MNRAS.417.2752R, 2011MNRAS.411..469R, 2013MNRAS.430.2249R, 2014MNRAS.445..414S, 2015MNRAS.451.3587R}.

According to the classification proposed by \citet{2018MNRAS.473.2160N}, biconical outflows can be divided into three types: symmetric, asymmetric, and nested, based on their intrinsic geometry of the outflows.
In symmetric bicones, double-peaked narrow emission lines with a stronger blue-shifted component may arise if additional effects such as dust obscuration or anisotropic radiation are considered. 
Asymmetric bicones can naturally produce such profiles when the receding cone (red) has a wider opening angle but lower brightness, while the approaching cone (blue) is more collimated and dominant. 
Nested bicones may result in two blue-shifted components, leading to an even more pronounced blue-dominated double-peaked profile.
\citet{2018MNRAS.473.2160N} propose that even moderate-luminosity AGNs are capable of driving significant outflows, potentially leading to feedback effects on galactic scales.
Therefore, outflows often cause a profile of core components along with blue wings \citep{2016ApJ...819..148K}, and the red wings are usually weak and may even not be seen, due to obscuration by galaxy disks or AGN dust tori \citep{2016ApJ...828...97B}.
More discussion of the outflows or rotating disks that trigger double-peaked profiles can be found in \citet{2010ApJ...715L..30L, 2011ApJ...735...48S, 2016ApJ...832...67N, 2019MNRAS.482.1889W, 2023A&A...670A..46M}.

In the case of \obj, as shown in panel (c) of Fig.~\ref{line}, the presence of clearly separated red-shifted and blue-shifted components, with the blue-shifted one being significantly stronger, is evident. 
The pronounced asymmetry of the double-peaked narrow [O{\sc iii}] emission lines corresponds well to the one-walled asymmetric bicone scenario (see their Figure 2) proposed by \citet{2018MNRAS.473.2160N}.
Meanwhile, nuclear activity can naturally account for the production of the observed jet structures, while the jets may be responsible for the highly shifted velocity of the broad H$\alpha$. 
The significant blue-shifted broad H$\alpha$ suggests a systematic motion of the BLR, possibly entrained by the jets. 
The presence of double-peaked profiles in part of the narrow lines indicates pronounced bidirectional motions within the NLR. 
Taken together with the global blue shifts observed in nearly all emission lines, these features strongly suggest that AGN-driven outflows are the dominant kinematic mechanism.

It is noteworthy that the multiple components of the [O{\sc iii}] doublet of \obj\ may correspond to different phases and/or stratifications of the outflowing gas. 
The red-shifted components are consistent with the NLR at the systemic redshift of the host galaxy, while the blue-shifted and extended (also significantly blue-shifted) components likely trace different outflowing gas components located in the blue cone. 
This interpretation is consistent with recent James Webb Space Telescope observations of extreme outflows in distant obscured AGNs, where gas has been found to exhibit multiple velocity components \citep{2024arXiv241202751Z, Zamora2025, 2025ApJ...989..230V}, without necessarily requiring a traditional biconical morphology.
This alternative interpretation does not contradict but rather complements the biconical outflow scenario.
However, further evidence will be required in future work to confirm these hypotheses.

In conclusion, the most plausible interpretation is that \obj\ is a Type-1.9 AGN, with spectral characteristics, including the double-peaked [O~{\sc iii}] and broad H$\alpha$, attributable to AGN-driven outflows.
An AGN-driven biconical outflow, which provides a natural explanation for the observed jet structures, the high shifted velocity of the broad H$\alpha$ component, and the pronounced asymmetry of the double-peaked narrow [O~{\sc iii}] emission lines.

Furthermore, the absence of broad H$\beta$, or the large flux ratio of broad H$\alpha$ to H$\beta$, observed in Type-1.9 AGNs is commonly attributed to significant obscuration of the central BLR and has been widely used in testing the unified model of AGNs. 
\citet{2024ApJ...961...82Z} employed two independent methods to estimate the black hole mass of SDSS J1241 + 2602 and found that its central BLR is unobscured, suggesting that Type-1.9 AGNs do not always conform to the traditional unified model of AGNs.
If the BLR is truly unobscured, the suppression of H$\beta$ would require an alternative physical explanation, such as a low-ionization state or high-density gas. 
Further studies can focus on determining whether \obj\ is indeed a Type-1.9 AGN and whether its BLR is obscured.

\section{Summary and Conclusions} \label{summary}

The discrepancy among previous studies concerning the presence of the double-peaked [O~{\sc iii}] doublet and broad Balmer lines in \obj\ was first noted. 
To address these issues, the optical spectrum of \obj\ obtained from SDSS DR16 was analyzed.
The stellar contributions from the host galaxy were modeled using 39 SSP templates, combined with a power-law component representing the AGN continuum. 
After subtracting both the host galaxy and AGN continuum emissions, Gaussian fitting was applied to the emission lines, with particular attention paid to the rest-frame wavelength ranges of 4800-5100\AA, 6200-6800\AA, 3650-4050\AA, and 4000-4500\AA.
The best-fitting results indicate that modeling each narrow line of the [O~{\sc iii}] doublet with a red-shifted and a blue-shifted component provides a better fit than using a single Gaussian, suggesting the presence of double-peaked profiles in the [O~{\sc iii}] doublet. 
Meanwhile, adding a broad Gaussian component around H$\alpha$ also provides a better fit.
Subsequently, a comparison between the SDSS and the SP2 spectra better confirmed the presence of a broad H$\alpha$ component, while no broad H$\beta$ component is detected in \obj. 
Taken together, these findings suggest that \obj\ is most plausibly classified as a Type-1.9 AGN, and the double-peaked narrow [O~{\sc iii}] profiles are more likely to be attributed to AGN-driven biconical outflows, rather than a dual AGN merger system or a rotating NLR.

\acknowledgments
We gratefully acknowledge the anonymous referee for giving us constructive comments and suggestions to greatly improve the manuscript.
Zhang gratefully thanks the kind financial support from GuangXi University and the kind grant support from NSFC-12173020 and NSFC-12373014, and the support from the Guangxi Talent Programme (Highland of Innovation Talents). 
Cheng \& Chen gratefully acknowledge the kind grant support from Innovation Project of Guangxi Graduate Education YCSW2024006. 
This manuscript has made use of the data from the SDSS projects (\url{https://www.sdss3.org}), managed by the Astrophysical Research Consortium for the Participating Institutions of the SDSS-III Collaborations. 
This research has made use of the NASA/IPAC Extragalactic Database (\url{https://ned.ipac.caltech.edu}) funded by the National Aeronautics and Space Administration and operated by the California Institute of Technology.

\bibliography{cpz}{}
\bibliographystyle{aasjournal}

\end{document}